# Location sharing without the central server.

## A peer-to-peer model for location sharing services

Dmitry Namiot
Lomonosov Moscow State University
Faculty of Computational Math and Cybernetics
Moscow, Russia
e-mail: dnamiot@gmail.com

*Abstract*— This paper describes a new model for sharing location info for mobile users. This approach can operate without the need for disclosing identity info to third party servers. It could be described as a safe location sharing model. The proposed approach creates a special form of distributed database that splits location info and identity information. In this distributed data store identity info is always saved locally. It eliminates one of the main concerns with location-based systems – privacy. This paper describes a model itself as well as its implementation in the form of HTML5 mobile web application.

## I. INTRODUCTION

It is a well-known fact that the question "where are you" is one of the most often asked during the communications. 600 billion text messages per year in the US ask, "Where are you?" (it is per Location Business Summit 2010 data [1]). A huge amount of mobile services is actually being built around this question so their main feature is user's location exchange.

Location, while being only one of the possible sensor readings of a modern smart phone, is probably the first attribute (candidate) to share for mobile users. Location information plays the major role in all context-aware applications [2]. The typical applications are well known and include for example geo-tagged context, friend-finder, recommendation systems, turn-by-turn navigation, etc.

In location-based service (LBS) scenarios we can describe the following actors [3]:

- Intended recipient, e.g., the service company, friends, parents, etc. This usually involves the use of a service provider that offers to forward your location to the intended recipient.

- Service provider, e.g., Google providing you with the Latitude application, or a restaurant recommendation system for near-by places. In contrast to the intended recipient, users usually do not have a primary goal of letting the service provider know their location – it is a by-product of getting a restaurant review or staying in touch with friends.

- Infrastructure provider, e.g., your mobile operator. While self-positioning systems such as GPS can work without an infrastructure provider, mobile phone users are often implicitly located in order to provide communication services (for example, route phone calls).

Some papers mentioned also so-called unintended recipients [3]. For example, we can mention accidental recipient, illegal recipient and law enforcement.

Interesting also, that in the most cases talking about LBS we assume that for a given system, the infrastructure provider needs to be trusted. In other words the need for sharing location data with infrastructure providers is non-discussable.

In the most cases location sharing is implemented as the ability for the mobile user (mobile phone owner) write down own location info in the some special place (special mobile application).

But it means of course, that user must be registered in this service (download some special application). And even more important – everyone who needs this information must use the same service too [1].

One of the biggest concerns for all location-based services is user's privacy. Despite the increased availability of these location-sharing applications, we have not yet seen wide adoption. It has been suggested that the reason for this lack of adoption may be users' privacy concerns regarding the sharing and use of their location information.

For example, the widely cited review of social networks practices [4] concluded that location information is preferably shared on a need to know basis, not broadcast. As it is mentioned in [5], current location sharing tools often present over-simplistic privacy settings by which users are forced to the binary alternative of sharing everything or nothing. Author suggests a special privacy-by-design approach to location sharing. This has led to the development of methods enhancing location privacy, and to the investigation of reasons for sharing location information. In the same time, some authors [6] explores human strategies and show that human strategies for de-anonymization and re-identification can be highly successful and thus pose a threat to location privacy comparable to computational attacks.

In the most location sharing surveys, participants were biased against sharing their location constantly, without explicit consent each time their location is requested. This suggests that people are cautious about sharing their location and need to be reassured that their private information is only being disclosed when necessary and is not readily available to everybody.

The key point for any existing service is some third party server that keeps identities and locations. We can vary the approaches for sharing (identity, locations) pairs but we could not remove the main part in privacy concerns – the third part server itself.

As mentioned in [7] peer opinion and technical achievements contribute most to whether or not participants thought they would continue to use a mobile location technology.

One possible solution is using peer-to-peer location sharing. The easiest way to apparently "solve" location privacy problems is to manually or automatically authorize (or not) the disclosure of location information to others. But we should see in the same time the other privacy issue that is not eliminated. Your location will be disclosed to (saved on) some third party server. For example, you can share location info in Google Latitude on "per friend" mode, but there is still some third party server (Google) that keeps your location and your identity

Typically we have now two models for location sharing in services. At the first hand is some formation of passive location monitoring and future access to the accumulated data trough some API. It is Google Latitude for example. Possible problems are privacy - some third party tool is constantly monitoring my location and what is more important – saves it on the some external server as well as the shorted life for handset's batteries.

Another model for location sharing is check-in. It could be an active (e.g. Foursquare), when user directly sets his/her current location or passive (e.g. Twitter) when location info could be added to the current message. A check-in is a simple way to keep tabs on where you've been, broadcast to your friends where you are, and discover more about other people in your community. But here we can see not only privacy issue - all my friends/followers can see my location but also a noise related issue – my location info could be actually interested only for the physical friends. For the majority of followers my location info (e.g. status from Foursquare in Twitter's time line) is just a noise [1]

Lets us describe some existing approaches in LBS development that targets the privacy.

One of the most popular methods for location privacy is obfuscation [8]. Obfuscating location information lowers its precision, e.g., showing only street or city level location instead of the actual coordinates, so that the visible (within our system) location does not correspond to the real one. For example, in Google Latitude we can allow some of the users get our own location info on the city level only. Sometimes even the random noise could be added to the real location data [9]. But once again – it is just a visible location. The central point (points) for such a system can have all the information.

Some articles prefer the using term spatial cloaking and describe it as the most commonly used privacy-enhancing technique in LBS. The basic idea of the spatial cloaking technique is to blur a user's exact location into a cloaked area that satisfies the user specified privacy requirements [10].

Another popular approach in the area of location privacy is "k-anonymity" [11]. As per this approach the actual location is substituted by a region containing at least $k - 1$ other users, thus ensuring that a particular request can only be attributed to "1 out of k" people. Of course, this approach has the disadvantage that if the region contains too few people, it has to be enlarged until it contains the right number of people. But in general k-anonymity protects identity information in a location-oriented context. In the same time, the group-composing algorithm is complex and the member peers are dynamic. The big question again is it core-level protection or just a view. In other words what kind of data do have inside of our system – anonymous location info right from the moment data being put into our system or it is just a view and data internally saved in raw formats.

Of course, the deployment of location privacy methods depends on the tasks our system is going to target. For example, obfuscating location information in case of emergency help system could not be a good idea. But from other side many geo-context aware applications (e.g. geo search) can use approximate location info.

Also we need to highlight the role of identity in LBS. It looks like combining identity with location info is just an attempt for delivering more targeted advertising rather then the need of the services themselves. It is obviously for example, that local search for some points of interests (e.g., café) should work for the anonymous users too.

Our idea of the signed geo messages service (geo mail, geo SMS) based on the adding user's location info to the standard messages like SMS or email. Just as a signature. So with this service for telling somebody 'where I am' it would be just enough to send him/her a message. And your partner does not need to use any additional service in order to get information about your location. All the needed information will be simply delivered to him as a part of the incoming message.

It is obviously peer-to-peer sharing and does not require any social network. And it does not require one central point for sharing location with by the way. Our location signature has got a form of the map with the marker at the shared location. And what is important here – the map itself has no information about the sender and recipient. That information exists only in the message itself. The map (marker) has no information about the creator for example. That is all about privacy [1].

More traditionally, peer-to-peer LBS refer to the way sharing information is traversed over the network [12]. For example, the P2P k-anonymity algorithm has several steps: select a central peer who will act as an agent for the group, next, the central peer will discover other k-1 different peers via single-hop or multi-hop to compose the group and finally find a cloaked region covering all locations that every peer may arrive.

In our article we are using "peer-to-peer" term at the first hand for highlighting the target party for the location-sharing request. It is "another peer" directly, rather than the central server (data store).

In terms of patterns for LBS this approach targets at the first hand such tasks as 'Friend finder' and the similar. In other words it is anything that could be linked to location monitoring.

The biggest danger of such systems is the recording of location information by service providers. Because every time a location update is shared, the service provider gets an update and is thus able to create detailed behavioral profiles of its customers (Google Latitude). As it is mentioned before, an ideal privacy-aware location sharing system should be able to share location information even without a central service provider receiving a copy of the entire movement track. It is exactly what Geo Message does.

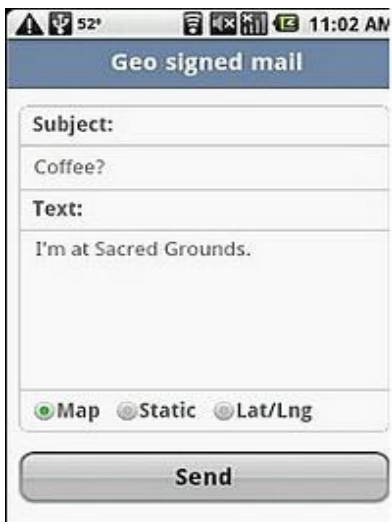

Figure 1. Geo Signed Mail

Geo Messages approach works and really eliminates identity revealing problems but it is pure peer-to-peer. What can we do if we need to monitor several participants simultaneously? It is simply not very convenient to jump from one message to another. Here we can offer a new peer-to-peer service that solves the privacy issues and lets you deal with several location feeds (location peers) simultaneously.

## II. OUR MODEL

What if we separate (split) the locations and identity? In other words rather than use one server that keeps all our data we will switch to some distributed architecture.

WATN (Where Are They Now) [13] requires no sign-in. It combines anonymous server-side data with local personalized records.

We can separate location info and identity data just in three steps:

a) assign to any participant some unique ID (just an ID, without any links to the personality)

b) save location data on the server with links to the above-mentioned IDs

c) keep the legend (descriptions for IDs, who is behind that ID) locally

In this case any participant may request location data for other participants from third party server (as per sharing rules, of course), get data with IDs and replace IDs (locally) with legend's data. With such replacement we can show location data in the "natural" form. For example: name (nick) plus location. And in the same time the server (third party server for our users) is not aware about names.

What does it mean technically?

Server keeps two things.

a) location info with meaningless IDs:

$ID_1$ -> (latitude, longitude)
$ID_2$ -> (latitude, longitude)
$ID_3$ -> (latitude, longitude)
Etc.

It is just a set of current geo-coordinates for users. Each user is presented via own ID.

and

b) social graph info – who is sharing location to whom. For example:

$ID_1$-> ($ID_2$, $ID_3$)
$ID_3$ -> ($ID_1$)
Etc.

Just a set of records states (as in example above) that user marked as ID1 shares location data with users ID2 and ID3

In the same time any local client keeps the own legend:

$ID_1$ -> (name or nick)
$ID_2$-> (name or nick)
Etc.

Note, that in this approach each client keeps own legend info. And because our clients are not aware about each other and there are no third party servers that know all registered clients. It means, obviously, that in this model the same ID may have different legends. Each client technically can assign own name (nick) for the same ID. Our social graph saves information (links between participants) using our meaningless IDs only. And the human readable interpretation for that graph can vary of course from client to client.

But that is probably very close to the real life, where the same person could be known under different names (nicks) in different contexts (e.g. compare some work environment and family space).

In general it is like keeping social graph, location and identity info in distributed database. But it is distributed on the server-client level, rather than on the traditional server-server level.

On practice, the structure could be a bit more elaborated. For example, in the current implementation we are saving the history - historical set of (latitude, longitude) pairs, we can keep some text messages associated with the current position etc. But it is just a set of features that does not changed the main idea – server-side store for anonymous location data and distributed client-side store with pseudo-personal data.

### III. WATN ALGORITHM

WATN has been implemented as mobile web application. HTML5 is significant there. Application uses W3C geo location [14] and local storage specification [15]. As per W3C documents HTML5 web storage is local data storage, web pages can store data within the user's browser.

Earlier, this was done with cookies. However, Web Storage is more secure and faster and our data is not included with every server request, but used only when asked for. It is also possible to store large amounts of data, without affecting the website's performance. The data is stored in key/value pairs, and web pages can only access data stored by them.

Storage is defined by the WhatWG Storage Interface as this:

*interface Storage {*
  *readonly attribute unsigned long length;*
  *[IndexGetter] DOMString key(in unsigned long index);*
  *[NameGetter] DOMString getItem(in DOMString key);*
  *[NameSetter] void setItem(in DOMString key, in DOMString data);*
  *[NameDeleter] void removeItem(in DOMString key);*
  *void clear();*
*};*

The DOM Storage mechanism is a means through which string key/value pairs can be securely stored and later retrieved for use. The goal of this addition is to provide a comprehensive means through which interactive applications can be built (including advanced abilities, such as being able to work "offline" for extended periods of time).

User agents must have a set of local storage areas, one for each origin. User agents should expire data from the local storage areas only for security reasons or when requested to do so by the user. User agents should always avoid deleting data while a script that could access that data is running.

Mozilla-based browsers, Internet Explorer 8+, Safari 4+ and Chrome all provide a working implementation of the DOM Storage specification.

We use local storage for saving legends for IDs as well as for the saving own ID for the each client.

As soon as the client calls the application we can restore his own ID from local storage (or obtains a new one from the server).

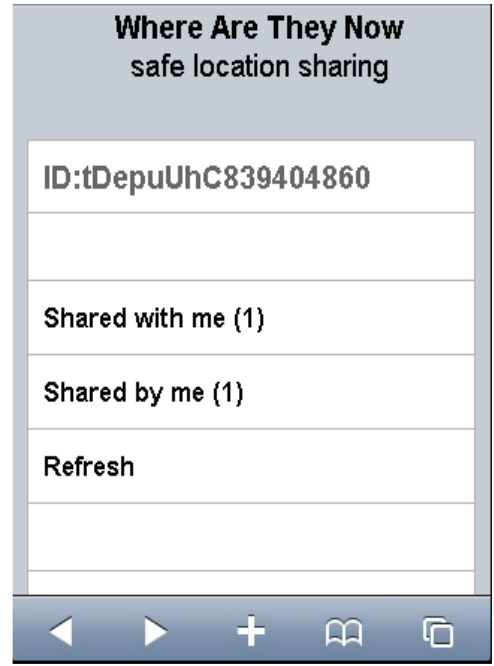

Figure 2.  WATN main screen

After that client saves location data on the server (it is check-in) and obtains shared location data (by the social graph). Server side part returns social graph data with ID's as JSON array. It is some like this:

[ {"id":$ID_1$, "lat":$lat_1$, "lng":$lng_1$},
  {"id":$ID_2$, "lat":$lat_2$, "lng":$lng_2$},
  …
]

For our server-side database it is just a plain select (no joins) where our own ID is a key. It is very important, because complex database queries in geo systems can seriously affect the performance.

After that we can simply match that array against the local database with identities. Client modifies received data and replaces IDs with known names from local database. So, after that our client side application is ready to show location data with names instead of IDs.

If our system is unaware about some legend, than of course it shows "raw" ID instead of name or nickname.

We can see (control) who is sharing location with us, as well as who can read our location info.

Note that using native JSON parsing and serialization methods provided by the browser, we can save the obtained data too. And technically it let us use the whole application in offline mode, playing with the last known data.

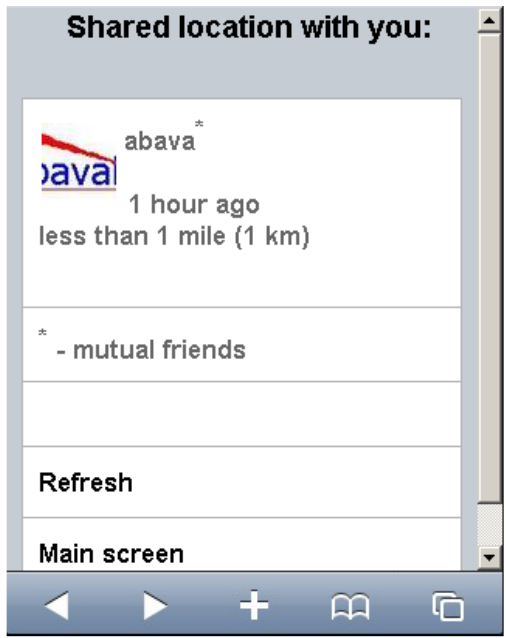

Figure 3. Shared location

And by the similar manner we can see to whom we share our own location info:

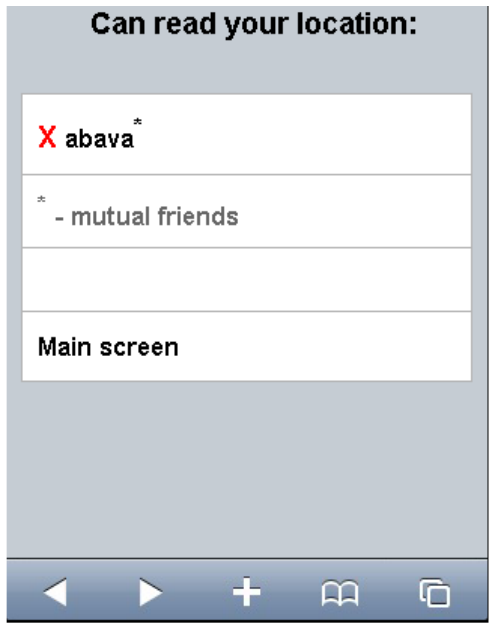

Figure 4. Share own location

And of course, we can cancel this link any time.

Where are the above-mentioned names (nicks) for IDs come from? WATN uses peer-to-peer sharing. It means that any user shares own location to another person directly. There are no circles, groups, lists etc. As soon as some user is going to show own location info to other person he simply sends notification about this to another email address (phone number in case of SMS). Actually the location could be shared to any person with known email address.

Such notification contains some text with explanation "what is it" and, what is obviously should be the main part of this process, a special link to WATN. This link contains an ID for the request originating party.

As soon as this link is fired, WATN application (client) becomes aware about two IDs: own $ID_1$ for this client (it is restored from the local storage – see description above) and $ID_2$ from the "shared with you" link (originated request ID). So, if notification is accepted, we can add social graph record (on the server) like

$$ID_2 \rightarrow ID_1$$

(client with $ID_2$ shares own location info with the client with $ID_1$. Or, what is technically equal, client with $ID_1$ identity may read location info for client with identity $ID_2$)

But because the notification link comes from some message (email or SMS), the receiver is aware about the context. Simply, he knows either email header ('*From*') or phone number or name in address book SMS comes from. It means, that based on that info, our receiver may assign some nick (name) for ID in "shared with you" link. Actually it is a part of confirmation: confirm and set some name. And that name (nick) we can save locally. So, it is like "two phase commit" in databases – save a new social graph record on the server and create a new legend (record for identity) locally.

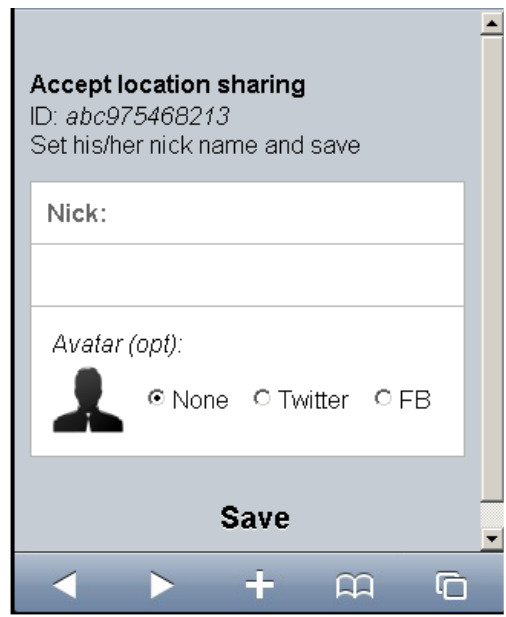

Figure 5. Accept location sharing request

And as a source for profile images (remember – there is no registration and profiles) we can use social networks (e.g. Twitter and Facebook). If you set a nick for new share that corresponds to Twitter's (Facebook's) account, the system can attach public photo from the social network.

Of course, as per above described scheme, the mutual location sharing could not be set automatically. The message with location sharing link is email (SMS) delivered outside of this application. So the application itself is completely unaware who is sending sharing message to whom.

It is obviously also, that in this schema each client has got own legends. We can have different names for the same ID (each client can technically assign own name)

Additional options include messaging and data clearing. As soon as you share your location info, you can leave messages attached to your location. WATN users that can read your location data will see your messages too.

And any time you can delete ID (as well as erase all the associated data) from the system. Note, that in case of any reconnection in the future WATN will assign a new ID for the user. There is no way to reuse some times once deployed ID. This feature also increases privacy. For the first time this model was proposed by D. Namiot and initial (alpha) implementation was presented in [16].

## IV. CONCLUSION

This article describes a new model for sharing location info in mobile networks. The proposed service WATN targets the privacy concerns for LBS applications users. It eliminates one of the biggest security-related problems in LBS applications: the need for third party server that keeps the most sensitive data – identities and locations for all users. WATN could be described as a safe location sharing.

Service is implemented as HTML5 mobile web application and is compatible with all the modern HTML5 mobile web browsers (iPhone, Android, Opera, Bada, etc.).
.

## REFERENCES

[1] D. Namiot Geo messages Ultra Modern Telecommunications and Control Systems and Workshops (ICUMT), 2010 International Congress pp. 14-19 DOI: 10.1109/ICUMT.2010.5676665

[2] D. Namiot and M. Schneps-Schneppe About location-aware mobile messages International Conference and Exhibition on. Next Generation Mobile Applications, Services and Technologies (NGMAST), 2011 14-16 Sept. 2011 pp. 48-53 doi: 10.1109/NGMAST.2011.19

[3] M. P. Scipioni and M. Lanngheinrich. I'm Here! Privacy Challenges in Mobile Location Sharing. Second International Workshop on Security and Privacy in Spontaneous Interaction and Mobile Phone Use (IWSSI/SPMU 2010), Helsinki, Finland, May 2010.

[4] D. Wagner, M. Lopez, A. Doria, I.Pavlyshak,V.Kostakos, I.Oakley, and T.Spiliotopoulos Hide and seek: location sharing practices with social media, Proceedings of the 12th international conference on Human computer interaction with mobile devices and services, September 07-10, 2010, Lisbon, Portugal

[5] M. Scipioni A privacy-by-design approach to location sharing UbiComp '12 Proceedings of the 2012 ACM Conference on Ubiquitous Computing pp. 580-583

[6] T. Fechner and C.Kray Attacking location privacy: exploring human strategies, UbiComp '12 Proceedings of the 2012 ACM Conference on Ubiquitous Computing, pp.95-98

[7] J.Tsai, P.Kelley, P.Drielsma, L.Cranor, J.Hong, and N.Sadeh Who's viewed you?: the impact of feedback in a mobile location-sharing application, Proceedings of the 27th international conference on Human factors in computing systems, April 04-09, 2009,

[8] M. Duckham and L.Kulik A formal model of obfuscation and negotiation for location privacy. In Proceedings of Pervasive 2005, pp. 152–170, Munich, Germany, 2005. Springer.

[9] J. Krumm Inference attacks on location tracks. In Proceedings of the Fifth International Conference on Pervasive Computing (Pervasive), volume 4480 of LNCS, pp. 127–143. Springer-Verlag, 2007

[10] M. Gruteser and D. Grunwald Anonymous usage of location-based services through spatial and temporal cloaking. In MobiSys '03: Proceedings of the 1st international conference on Mobile systems, applications and services, pp. 31–42, New York, NY, USA, 2003. ACM

[11] A. Paiva, E.Monteiro, J. Rocha, C. Baptista, and A.Silva "Location Information Management in LBS Applications", Encyclopedia of Information Science and Technology, Second Edition, pp. 2450-2455, 2009

[12] J. Xu, M.Xu, and N. Zheng Mobile-Aware Anonymous Peer Selecting Algorithm for Enhancing Privacy and Connectivity in Location-Based Service e-Business Engineering (ICEBE), 2010 IEEE 7th International Conference on Nov. 2010 pp. 172 – 177 DOI: 10.1109/ICEBE.2010.32

[13] WATN service http://watn.servletsuite.com Retrieved: Oct 2012

[14] Geolocation API Specification (2010, Sep. 7) http://www.w3.org/TR/geolocation-API/

[15] M. Casario, P.Elst, C.Brown, N.Wormser, and C. Hanquez HTML5 Solutions: Essential Techniques for HTML5 Developers 2011, pp. 281-303, DOI: 10.1007/978-1-4302-3387-9_11

[16] D. Namiot and M. Sneps-Sneppe Where Are They Now – Safe Location Sharing. A New Model for Location Sharing Services, Internet of Things, Smart Spaces, and Next Generation Networking, Lecture Notes in Computer Science, 2012, Volume 7469/2012, pp. 63-74, DOI: 10.1007/978-3-642-32686-8_6